\documentclass[aps,showpacs]{revtex4}
\usepackage{graphicx}
\usepackage{epsfig}
\begin{document}
\title{A theoretical description of energy spectra and two-neutron
separation energies for neutron-rich zirconium isotopes}
\author{J.E.~Garc\'{\i}a--Ramos$^1$,K.~Heyde$^2$,R. Fossion$^2$,
V. Hellemans$^2$ and S. De Baerdemacker$^2$}
\affiliation 
{$^1$Departamento de F\'{\i}sica Aplicada, Universidad de Huelva,
21071 Huelva, Spain.\\
$^2$Institute for Theoretical Physics,Vakgroep Subatomaire en Stralingsfysica,
Proeftuinstraat 86, B-9000 Gent, Belgium.\\}

\begin{abstract}
{Very recently the atomic masses of neutron-rich Zr isotopes, from
$^{96}$Zr to $^{104}$Zr, have been measured with high precision. 
Using a schematic Interacting Boson Model (IBM) Hamiltonian, 
the evolution from spherical to 
deformed shapes along the chain of Zr isotopes, 
describing at the same time the excitation energies
as well as the two-neutron separation energies, can be rather well reproduced. 
The interplay between phase transitions and configuration mixing of
intruder excitations 
in this mass region is succinctly addressed.}
\end{abstract}
\maketitle

Nuclear masses and binding energies, or more in particular two-neutron
separation energies (S$_{2n}$), form a very important observable that
characterize a given nucleus and provide information about nuclear
correlations providing at the same time a stringent test for nuclear
models.

The goal of this paper is to make use of a technique developed in
Ref.~\cite{Foss02} in order to describe at the same time the energy
spectra and S$_{2n}$ values obtained from a very recent experimental
study of masses for the neutron-rich Zr isotopes (see refs.
Ref.~\cite{Rint04,Audi03}) and nearby Sr and Mo isotopes
\cite{Hage05}. This transitional region is of particular interest
because a rapid change in the structure of Zr isotopes from mass A=98
and onwards is observed,{\it i.e.}~a rather sudden change from
spherical to well deformed shapes \cite{mol81,Heyd84,woo99}.  This
region is known for the appearance of deformed intruder states that
even become the ground state and initiate the onset of a region of
deformed nuclei for the heavier Zr isotopes \cite{Heyd87} (beyond
N=58).  Therefore, a study has been attempted in order to analyze the
experimental variation observed in S$_{2n}$ values.

For a theoretical description of the Zr isotopes, encompassing both
the low-lying excitations as well as the intruder states, one should
be using a very large shell-model configuration space using the
corresponding effective nucleon-nucleon interaction
\cite{Caurier05}. An attempt in that respect has been carried out for
$^{92}$Zr \cite{Fran05} using a restricted model space considering
both proton and neuron orbitals outside of a $^{88}$Sr core nucleus.
In view of the fact that besides neutrons filling the 50-82 shell one
would need to consider protons in the 28-40 shell, including proton
excitations into the $1g_{9/2}$ configuration, calculations would
become unfeasible. Therefore, we start from a strongly truncated model
space, however keeping the pairing and quadrupole force components
within the Interacting Boson Model (IBM) approximation\cite{Iach87}.
This model approximates the interacting many-fermion problem using as
the major degrees of freedom, $N$ pairs of valence nucleons that are
treated as bosons, carrying angular momentum either $0$ (the $s$
bosons) or $2$ (the $d$ bosons). This model is very appropriate in
order to describe even-even medium-mass and heavy nuclei and
transitional nuclei. Even here, treating proton and neutron bosons
explicitly, one risks to be involved with too many model parameters.
Therefore, in the present description of the Zr isotopes, we make use
of an approach in which we restrict to the use of identical
bosons. This act of truncation naturally implies that one has to
replace the Hamiltonian by an effective IBM Hamiltonian describing the
interactions amongst these identical bosons \cite{Heyde04}.  Our
approach here is very similar to a recent study of the Pt nuclei
\cite{Mccu05}, a region in which there exist clear indications of the
presence of intruder excitations.

The IBM Hamiltonian used is composed from a single-boson energy term, a quadrupole and an 
angular momentum term, 
\begin{equation} 
\hat H= \varepsilon_d \hat n_d -\kappa \hat Q\cdot \hat Q +\kappa' \hat L
\cdot \hat L,
\label{h-cqf}
\end{equation}  
where $\hat n_d$ denotes the $d$ boson number operator and
\begin{equation}
\hat L= \sqrt{10} (d^\dag \times \tilde d)^{(1)} ,
\end{equation} 
\begin{equation}
\hat Q = s^\dag \tilde d + d^\dag \tilde s + 
\chi (d^\dag \times \tilde d)^{(2)} .
\end{equation} 
We point out that in more realistic
calculations, the values  $\varepsilon_d>0$ and  $\kappa>0$ \cite{Scholten78,
Scholten80,Stachel82,Casten78,Chou97} have
been used. Moreover, the $E2$-transition operator is taken to have the same
structure of the quadrupole operator $\hat Q$ appearing in the Hamiltonian, 
\begin{equation}
\hat Q(E2) = e_{eff}\hat Q.
\end{equation}
This approach is known as the Consistent-Q Formalism (CQF) \cite{Cast88}.

The definition of the two-neutron separation energies is the
following (starting from the binding energies):
\begin{equation} 
\label{sn2-def1}
S_{2n}=BE(N)-BE(N-1),
\end{equation} 
where $N$ denotes the number of valence nucleon pairs and it is assumed
that we are treating nuclei belonging to the first half of the neutron shell
$(50-82)$ filling up with increasing mass number.

The Hamiltonian (\ref{h-cqf}) generates the energy spectrum of each
individual Zr nucleus and will be called ``local Hamiltonian''. For
the correct description of binding energies, one needs to add that
part of the Hamiltonian that does not affect the spectrum and that
will be called ``global Hamiltonian''(and so depends on the total
number of bosons only \cite{Foss02}). The simplest interpretation of
the IBM global part comes from the fact that this part describes the
overall smooth varying energy term and can be associated with the
structure of the Liquid Drop Model. Therefore, the description we use,
within the context of the IBM, is somehow similar to the Strutinsky
method \cite{Stru67,Stru68} in the sense that the global part takes
care of the major contribution to the nuclear binding energy BE, while
the local part, that is notably smaller, modulates this global
behavior and describes the local correlations.   An alternative
way to incorportate binding energy effects in the IBM results from
explicitly taking into account the s-boson one-body contribution
$\varepsilon_{s} \hat{n}_{s}$ and the s-boson two-body interaction
energy $u_0 s^\dag s^\dag ss$. 
One can, however, eliminate these s-boson terms taking into account
the conservation of the number of bosons, giving rise to
terms proportional to  $N$ and $N(N-1)$ in the binding
energies while proportionals to $N$ in the two neutron separation
energies (see Ref.~\cite{Iach87-12}). Thus this procedure 
is equivalent to the method used here and in Ref.~\cite{Foss02}.

The contribution of the global part of the Hamiltonian to the S$_{2n}$
is expressed as
\begin{equation}
S^{gl}_{2n}={\cal A} +{\cal B} N,
\label{s2n-gl}
\end{equation}
where ${\cal A}$ and ${\cal B}$ are assumed to be constant along a
chain of isotopes (see \cite{Foss02}). Therefore,
\begin{equation}
S_{2n}(N)={\cal A} +{\cal B} N+BE^{lo}(N)-BE^{lo}(N-1),
\end{equation}
where $BE^{lo}$ is the local binding energy derived from the Hamiltonian 
(\ref{h-cqf}).

In order to obtain a global description of the Zr isotopes, we prefer to consider a reduced
number of parameters, keeping the Hamiltonian (\ref{h-cqf}) as constant as possible  when 
passing from isotope to isotope. Because the
heavier Zr isotopes exhibit energy spectra that are the more rotational, the ratios
$E(4^+_1)/E(2^+_1)$ and $E(6^+_1)/E(2^+_1)$ are used for fixing the
parameters of the Hamiltonian, while in the lighter and medium
isotopes the
energies of the $2_1^+$, $2_2^+$ and $0_2^+$ states are used to fix the
Hamiltonian. 
The fact that, in particular, for the $4_1^+$ and $0_2^+$ states,
large deviations appear when comparing the calculated $B(E2)$ values
with the experimental data (see also Table \ref{table-be2}), points
out that large
contributions of configurations outside of the purely collective IBM
model space are present in those states. This effect is enhanced
for the low number of bosons present in the lighter isotopes.
Thus, even though we cannot
describe nuclear excited state properties in detail for the light Zr
nuclei, what is, however, more relevant in the present study is the
important change in the low-lying energy spectra when passing from
mass $A=98$ towards mass $A=100$ that is  reproduced rather
well. This is also essential in deriving the binding energy effects
and the $S_{2n}$ values.

\begin{center}
\begin{figure}[hbt]
\includegraphics[height=13cm]{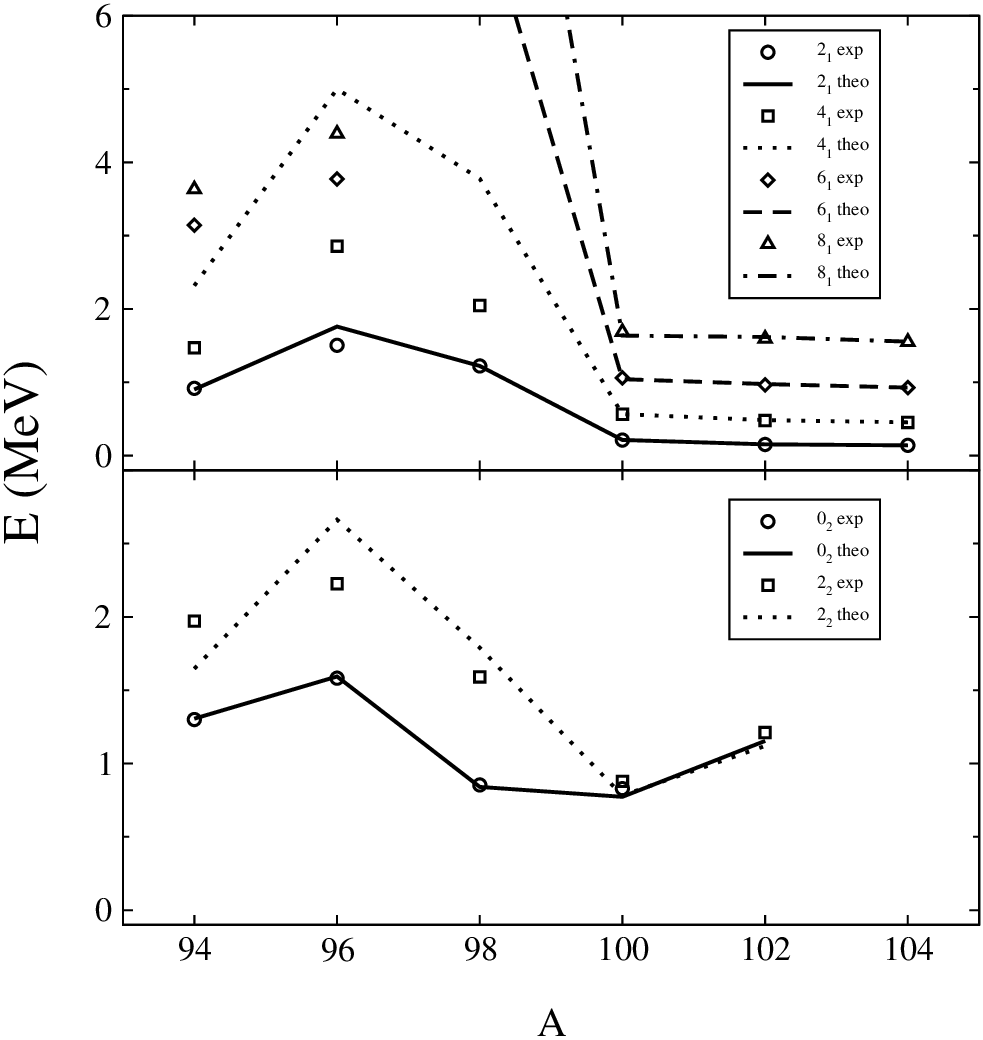}
\caption{Comparison between the theoretical and experimental energy
levels (Refs. \cite{Tuli92}-\cite{Blac91}) for the neutron-rich Zr isotopes.}
\label{fig-spectra}
\end{figure}
\end{center}

The experimental data for the different Zr isotopes have been
taken from references \cite{Tuli92}-\cite{Pante05}. 
Note that in this region, one expects the presence of intruder states
but, as discussed before, those excitations are outside of the model
space and are absorbed within an effective way within the changing
parameters of the Hamiltonian (\ref{h-cqf}) we are using in the
present study.  Therefore, the intruder states should be identified
but not to be considered explicitly in the fitting procedure; 
this is the main difference with the technique used in
Ref.~\cite{Mccu05}, where the intruder states are considered in the
fitting procedure.  The more clear-cut examples, in this region,
where extra configuration and thus mixing will appear are $^{98}$Zr
and $^{100}$Zr \cite{Heyd84,Heyd88,Cost99}. In $^{98}$Zr the $0_3^+$
state at $1.436$ MeV is supposed to form the head of an intruder band
\cite{Lher94}. This then could result in mixing between the $4^+$
states resulting in pushing down the $4^+_1$ state, which energy is
indeed overestimated by the IBM calculation (see figure
\ref{fig-spectra}). In the case of $^{100}$Zr, the $0_2^+$ state at
$0.331$ MeV is considered as an intruder state while the $0_3^+$ state
at $0.829$ MeV is considered to be the regular state.  Therefore, in
figure \ref{fig-spectra}, the $0^+$ state that is plotted is indeed
the state $0_3^+$. Information on the characteristics of the excited
$0^+$ states in these nuclei is gained by studying $E0$ properties as
discussed in detail by Wood et al.  \cite{woo99}. Note that in figure
\ref{fig-spectra} the theoretical states for the lighter isotopes with
angular momenta $6$ and $8$ stay out of scale, which shows again the
influence of the intruder states in the lower part of the spectra.  An
additional problem in the description of this region arises from the
low number of bosons we should use, in other words because of the
proximity of the shell closure for neutrons. That creates two
inconveniences, on one hand it is difficult to give a reasonable
description for high angular momentum (note that the maximum angular
momentum one can construct coupling IBM bosons is the double of the
number of bosons) as observed recently in $^{96}$Sr and $^{98}$Zr
nuclei \cite{Wu04} and, on the other hand, in the spectrum there
appear non collective excitation that can only be described by a shell
model calculation (see e.g. \cite{Fran05}).
 
There is a very poor knowledge of $E2$ transition rates in this mass region
and so it is difficult to fix the value of parameter $\chi$ because its value
it is not well defined using the information about energy spectra only.
Therefore its value is fixed taking
into account the information from other calculations addressing nuclei in
nearby mass regions \cite{Chou97}.  

In the present calculations, we count bosons starting from a
$Z=40, N=50$ core $^{90}$Zr which has a rather high-lying first excited
state. This is at variance with recent shell-model calculations for
this mass region \cite{Fran05} that start from a $^{88}_{38}$Sr$_{50}$
core. Since we make use of the interacting boson model approach (IBM)
that does not discriminate between proton and neutron bosons,
collective properties are mainly governed by the symmetry structure of
the IBM Hamiltonian and the total number of bosons present.  Using
proton and neutron boson degrees of freedom, in a more detailed
IBM-study, one needs both proton and neutron bosons to be active in
order for collective deformation effects to appear. This then would
imply a different choice of a closed proton core, conform with the
shell-model.
             
The parameters of the Hamiltonian are summarized in Table
\ref{table-par}. Here, one notices that $\chi$ is fixed at the value
of $-0.8$.  The values of $\kappa$ are restricted to $0.032$ MeV for
the Zr nuclei situated in the spherical and transitional region and to
$0.046$ MeV for the Zr nuclei exhibiting rotational-like energy
spectra in the ground band. Note the high value of $\varepsilon_d$ in
the case of $^{96}$Zr which is due to the abnormally high excitation
energy of the $0_2^+$ state that is described in terms of a subshell
closure at the neutron $N=56$.

We present in Table \ref{table-be2} the comparison between the
experimental and theoretical $E2$ transition rates, keeping $\chi$ =
-0.8 and fixing $e_{eff}$ = 0.159 e~b for reproducing the $B(E2;2^+_1
\rightarrow 0^+_1)$ value in $^{100}$Zr. The agreement of the results
is reasonable except for the transitions $4^+_1 \rightarrow 2^+_1$ and
$0^+_2 \rightarrow 2^+_1$ in $^{94}$Zr, which suggests that those
states are outside of the IBM collective space to a large extent, as
is also corroborated by the shell-model calculations carried out in
$^{92}$Zr \cite {Fran05}. The $4^+_3$ located at $2.330$ MeV is a good
candidate as IBM partner, while there is no other candidate for $0^+$.

Once the energy spectra for the even-even Zr isotopes have been fitted,
one derives the global part of S$_{2n}$ (\ref{s2n-gl}) assuming the equivalence,
\begin{equation} 
\label{s2n-lin}
S_{2n}^{gl}\equiv {\cal A}+{\cal B} N=S_{2n}^{exp}-S_{2n}^{lo}.
\end{equation} 
Note that left hand side of equation (\ref{s2n-lin}) can be written in
terms of the atomic number, $A$, through a trivial transformation in
the parameters ${\cal A}$ and ${\cal B}$.
In practice, the right-hand side of Eq.~(\ref{s2n-lin}) is not an exact 
relation but results approximately in a straight line. As a
consequence the linear part is derived from a best fit to the data points, 
obtained when plotting the right-hand side of (\ref{s2n-lin}). 
We like to stress at this point (see also Ref.~\cite{Foss02})
that both the results concerning relative energies (energy spectra)
and the energy surfaces do not depend on the values of ${\cal A}$ and
${\cal B}$ as determined here.

\begin{center}
\begin{figure}[hbt]
\includegraphics[height=11cm]{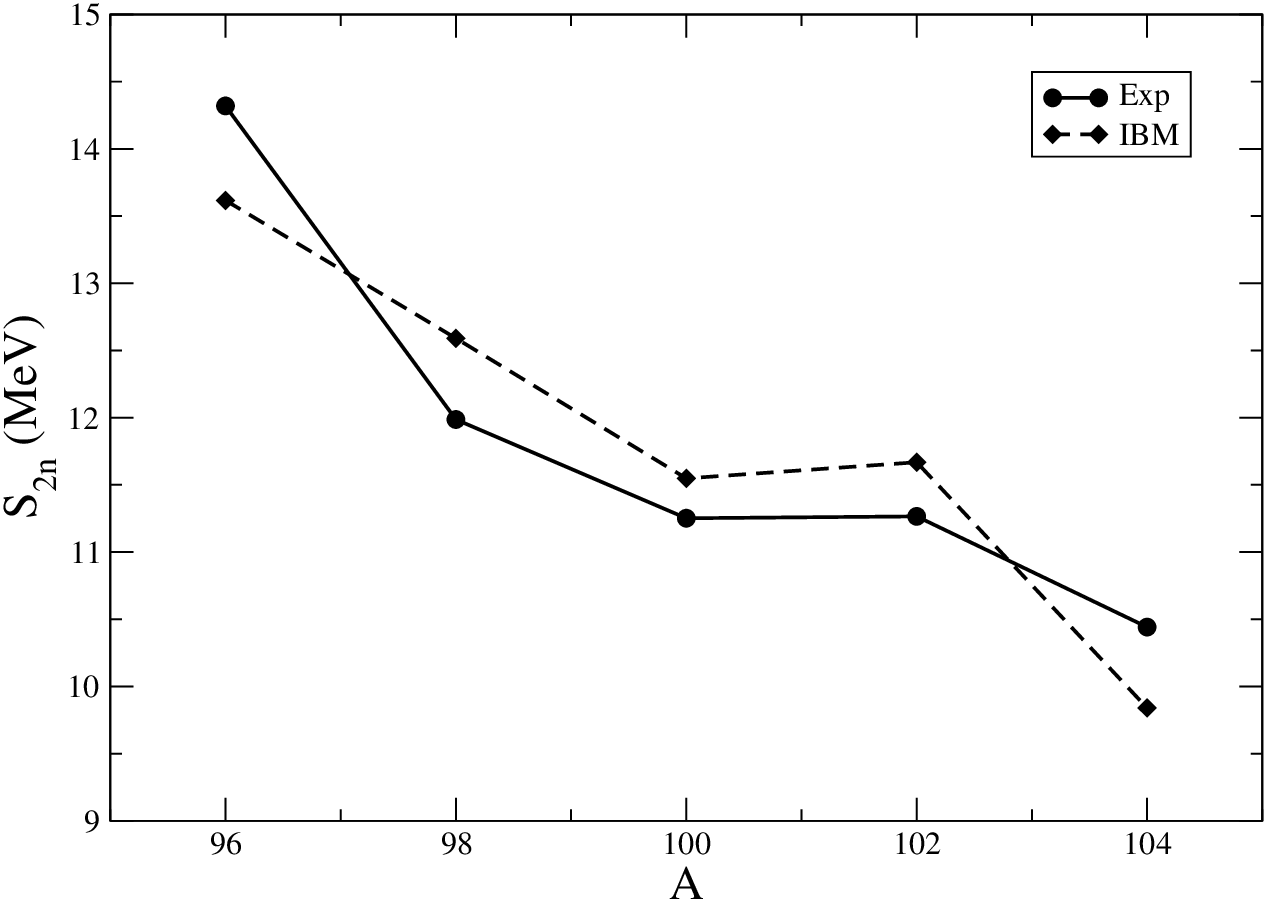}
\caption{Two-neutron separation energies for neutron-rich Zr
isotopes. Full lines correspond to experimental data \cite{Rint04}, while dashed
lines correspond to theoretical calculations.}
\label{fig-s2n}
\end{figure}
\end{center}

The global part of S$_{2n}$ corresponds to the values ${\cal A}=67.4$
MeV and ${\cal B}=-0.946$ MeV (using the atomic number, $A$, as
variable). The comparison between the experimental data and the
theoretical results, combining the global and the local part, is given
in figure \ref{fig-s2n}. We like to point out that the approximately
flat behavior of S$_{2n}$ at A=100,102 is rather well reproduced and
corresponds precisely to the neutron number where the energy spectra
rapidly change from spherical into deformed structures.

\begin{center}
\begin{figure}[hbt]
\includegraphics[height=11cm]{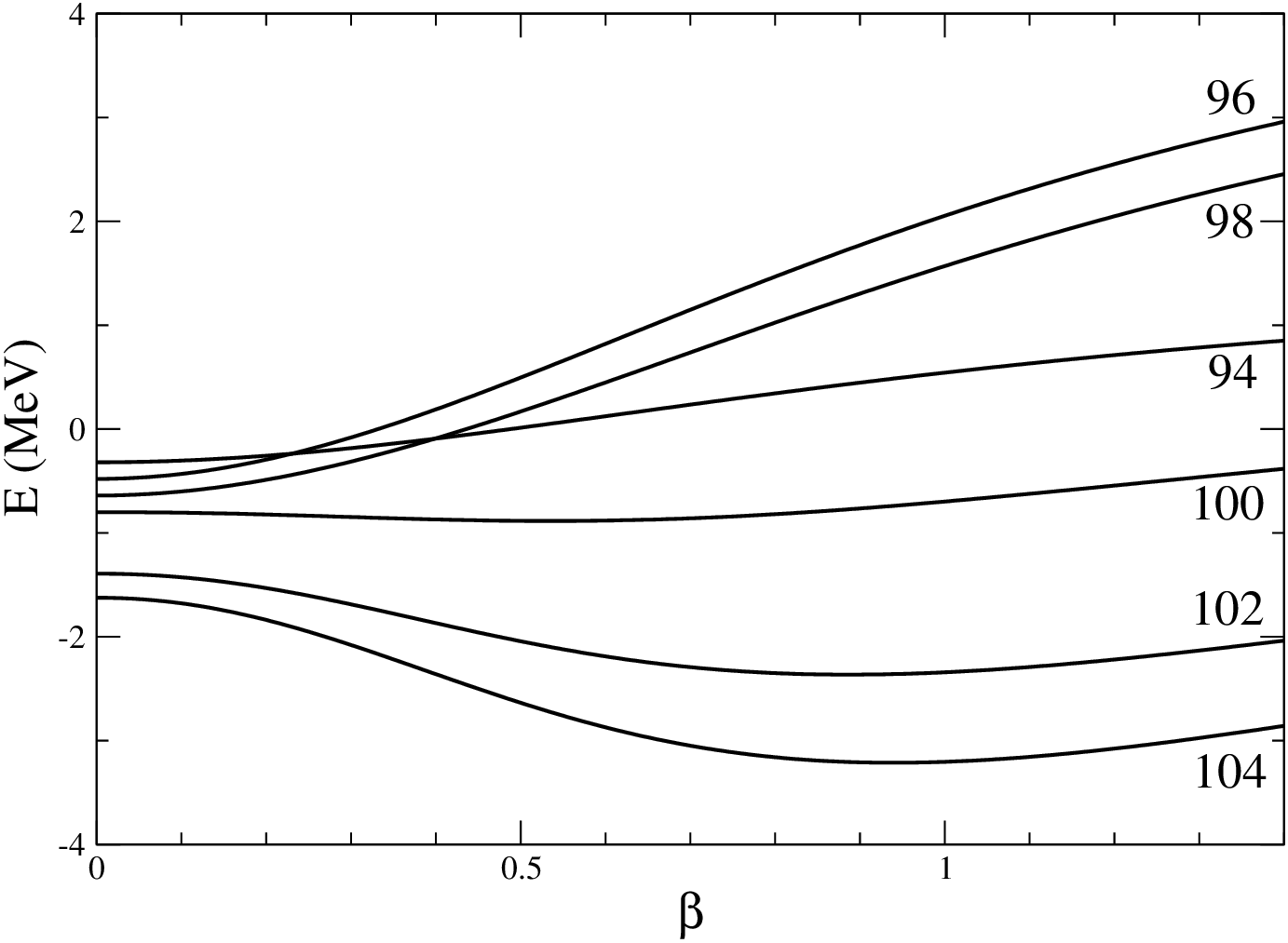}
\caption{Energy surfaces for Zr isotopes using the IBM intrinsic
state formalism. The number on each curve denotes the atomic mass
number.}
\label{fig-surf}
\end{figure}
\end{center}

Extra information that can be obtained from the present IBM calculation is
the energy surface as a function of the deformation parameters. This can most
easily be studied using the intrinsic-frame formalism. Here, the interacting 
many-boson problem is solved defining a new kind of boson -dressed bosons-
that is built as a linear combination of $s$ and $d$ bosons and constructing a 
trial wave function as a condensate of $N$ such bosons \cite{Diep80b,Gino80}. 
The problem is solved by minimizing the expectation value of the Hamiltonian with 
respect to the deformation parameters, that reduces to 
one, $\beta$. In figure \ref{fig-surf}, we present the energy
surfaces for the different Zr isotopes according to the parametrization as described 
in Table \ref{table-par}. It is clearly observed
that $^{94-98}$Zr remain spherical. Note that the flat energy surface of $^{94}$Zr 
appears due to the low number of bosons
for this nucleus, N=2, and to the fact that the depth of the potential energy 
surface is proportional to $N(N-1)$. $^{100}$Zr is a special case because this 
nucleus appears to be situated very close to the critical region where two minima 
coexist \cite{Iac00},one spherical and one deformed. One notices that the energy 
surface is very flat which is caused by the fact that a deformed minimum and a 
spherical maximum appear that are almost degenerate. This indicates that one is close 
to the critical area  \cite{Iac00}
Finally, $^{102-104}$Zr become well deformed.
Recent theoretical studies using relativistic mean-field (RMF) methods
\cite{Hema04} (concentrating on nuclear charge radii, mainly) and Hartree-
Fock-Bogoliubov (HFB) methods \cite{Blaz05} (studying the Zr isotopic chain
up to the two-neutron drip line) have concentrated on shape properties and 
their variation from spherical towards strongly deformed Zr nuclei.
In this brief report, we have studied 
the energy spectra and the S$_{2n}$ values for the neutron-rich even-even Zr isotopes 
in a consistent way using the IBM framework. The new experimental data on masses are rather 
well reproduced starting from a schematic calculation. 
From inspecting the corresponding energy surface diagrams, one clearly observes
how the Zr isotopes evolve from spherical into deformed shapes passing
through a region where two minima exist. 

The calculations that have been carried out imply the possible
presence of a phase transition, being $^{98}$Zr an almost critical
nucleus. On the other hand, this mass region can also be described
using configuration mixing (this will be shown elsewhere) in such a
way that for $^{98}$Zr, regular and intruder states coexist very
closely in energy although it should be necessary to see if the
S$_{2n}$ values can be appropriately described.  Both descriptions can
seem equivalent, but there exist clear differences \cite{Heyd04} in
the size of the model spaces. In the calculations carried out here,
the model space only consists of states with $N$ bosons and we can
follow the sequence of all states within this space as a function of
boson number and a smoothly changing Hamiltonian (see
Eq.~(\ref{h-cqf}) and Table \ref{table-par}).  On the other hand, when
treating the presence of intruder states explicitly, one has to expand
the configuration space such that both $N$ and $N+2$ boson
configurations are considered. It might be that the occurrence of the
deformed states as lowest states, forming the ground-state band from
$^{100}$Zr and onwards, can be associated with an adiabatic crossing
of the more deformed ($N+2$ boson configurations) and the more
spherical ($N$ boson configurations) \cite{Heyd87}. A microscopic
origin can then be related with the possibility of exciting protons
from the $2p_{1/2}$ into the $1g_{9/2}$ orbital and the subsequent
large proton-neutron interaction energy gain with the neutron
$1g_{7/2},1h_{11/2}$ orbitals beyond N=58
\cite{Fed77,Fed79,Etche89,Kir93}.  Clearly, more work needs to be
carried out in order to see if there exist mass regions where phase
transitions are induced by the presence of low-lying intruding
configurations and the corresponding configuration mixing.

The authors are grateful to J.\"{A}yst\"{o} for interesting
discussions. This work have been partially supported by the Spanish
DGI under project number FPA2003-05958. Financial support from the
``FWO-Vlaanderen'' (K.H and V.H.), the University of Gent (S.D.B and
K.H.) and the IWT (R.F.) as well as from the OSTC(Grant IUAP \# P5/07)
is also acknowledged.

\newpage
\begin{table}
\begin{tabular}{ccccccc}
\hline
A & 94 & 96 & 98 & 100 & 102 & 104 \\
N &  2 &  3 &  4 &   5 &   6 &   7 \\
\hline
$\varepsilon_d$  & 0.550 & 0.750 & 0.380 & 0.311 & 0.211 & 0.213 \\
$\kappa$      & 0.032 & 0.032 & 0.032 & 0.032 & 0.046 & 0.046 \\
$\chi $       & -0.8  & -0.8  & -0.8  & -0.8  & -0.8  & -0.8  \\
$\kappa'$     & 0.05  & 0.17  & 0.15  & 0     & 0     & 0     \\    
\hline
\end{tabular}
\caption{Parameters, describing the IBM Hamiltonian of
  Eq.~(\ref{h-cqf}), for the Zr isotopes.} 
\label{table-par}
\end{table}

\begin{table}
\begin{tabular}{cccc}
\hline
Isotope & Transition&B(E2) ($e^2b^2$) Exp. & B(E2) ($e^2b^2$) Theo.\\
\hline
$^{94}$Zr  & $2_1^+\rightarrow 0_1^+$ & 0.013   & 0.053\\
$^{94}$Zr  & $4_1^+\rightarrow 2_1^+$ & 0.002   & 0.027\\
$^{94}$Zr  & $0_2^+\rightarrow 2_1^+$ & 0.370   & 0.037\\
$^{96}$Zr  & $2_1^+\rightarrow 0_1^+$ & 0.010   & 0.088\\
$^{100}$Zr & $2_1^+\rightarrow 0_1^+$ & 0.226   & 0.226\\
$^{100}$Zr & $8_1^+\rightarrow 6_1^+$ & 0.336   & 0.251\\
$^{102}$Zr & $2_1^+\rightarrow 0_1^+$ & 0.297   & 0.347\\
\hline
\end{tabular}
\caption{Comparison between experimental and theoretical B(E2)
  values. The parameters of the quadrupole operator are
  $e_{eff}=0.159$ eb and $\chi=-0.8$.}

\label{table-be2}
\end{table}

\end{document}